\documentclass{article}
\usepackage{graphicx} 

\title{Stability of Zipf’s law for cities and occult spatial dependence effects: A study on the OECD countries}
\author{Rolf Bergs \\ PRAC Bergs and Issa Partnership Co.\and Uwe Neumann \\ RWI - Leibniz-Institute for Economic Research}

\date{February 2025}

\begin{document}

\maketitle
\begin{abstract}
We investigate spatial dependence in Zipf's law for cities among the OECD countries. The aim is to identify an upper tail of the distribution that follows a power law (Pareto) but is perturbed by spatial autocorrelation, as indicated by a coefficient with a significant minor or major deviation from a distribution corresponding to a (non-spatial) Zipf law. For that purpose, we augment the usual Pareto model with a spatial weight matrix and apply SEM/SAR regressions. The results for the OECD countries are mixed. In particular, we investigate the rank-size distribution of cities by estimating local Moran-I coefficients (LISA) along the city ranks to locate the causes of spatial dependence. As an example, we demonstrate the approach for Belgium, a medium-sized OECD country.
\end{abstract}
\section{Introduction}
This is a report on preliminary results of an extended investigation of spatial dependence in Zipf's law for cities. We investigated a large 2024 urban dataset (World Cities Database) provided by NGIA, US Geological Survey, US Census Bureau, and NASA to explore a specific aspect of Zipf's law for cities. The purpose of our research is to further substantiate insight of a previous article on spatial dependence in Zipf’s law that dealt with the simulation of randomly generated artificial distributions and four real world datasets covering the USA, Germany, UK, and Slovenia (Bergs 2021). That analysis could markedly prove the potential existence of major spatial dependence and thus confirms a bias of regression coefficients. However, perturbations from spatial lag or spatial error on the Pareto coefficient based on real world data at the country level have shown to be partly zero or - if at all - modest, even though partly significant at the 0.01 level, hence "weak but not negligible". The relevance of spatial dependence in Zipf’s law was later substantiated from the viewpoint of urban scaling (Arcaute and Ramasco 2021; Xiao and Gong 2022; Griffith 2022). A spatial dependence effect on the rank-size distribution of cities and the Zipf law suggests that this famous power law is not simply a statistical tautology, as Gan at al (2006) postulated, but in fact may indicate a major relevance for the functional interaction between urban and rural regions. While with respect to the earlier applications of Zipf's law, which derives from linguistic studies, other interdependencies could affect statistical results (Zipf 1932), spatial effects cannot be ruled out for urban systems. Empirically, it could be shown that there is a close relationship between the overall centrality of cities (i.e. the economic role of any city in a national urban system) and the same cities with respect to their urban and rural neighborhood (e.g. Hsu 2011). This can potentially activate local spatial autocorrelation among either proximate or distant city ranks of a country. Spatial autocorrelation can thus be significantly positive or negative or - of course - just insignificant. 
\\
\hspace*{3em} To our knowledge, to date there have been no studies that explore distance effects on the rank-size distribution in urban systems for a broader spectrum of countries. Therefore, it was our aim to do just that and to test spatial dependence for a larger range of countries by addressing the OECD group. For long, the OECD has represented the countries of the core industrialized Western world, but it now also includes countries that thirty years ago still belonged to the developing world like for instance Colombia, Costa Rica, Chile, Mexico, Korea and Turkey. The OECD thus covers countries from major world regions including the Americas, East Asia, Oceania, and Europe.
\section{Methodology}
Methodologically, we apply a SEM/SAR regression analysis:

\begin{equation}
    \centering
    ln(R)=\rho W ln(R)+ln(C)-\alpha ln(S)+\epsilon
    \end{equation}

\begin{equation}
    ln(R)=ln(C)-\alpha ln(S)+\nu
 \end{equation}
 \begin{equation}
     \nu=\lambda W\nu+\epsilon
 \end{equation}

The three equations, SAR defined by (1) and SEM by (2) and (3), show augmented Pareto regressions with the usual notation of spatial econometrics, where in our model R means rank and S means population; C is a constant. The variable R is defined as "rank-1/2" to avoid a standard error implied bias (Gabaix and Ibragimov 2011) and estimated by maximum likelihood (ML). Even though, the Gabaix-Ibragimov approach is specifically aimed to reduce the bias of an OLS estimate for small samples, we follow Le Gallo and Chasco (2006) who found more robust ML estimates with "rank-1/2" as well. More importantly, in our study the differentiation of the distribution takes account of a vital former debate about the city rank-size distribution being either Pareto- or log-normally distributed or hybrid (Malevergne et al. 2011). 

\hspace*{3em}The SEM/SAR procedures were carried out for (i) the entire distribution per country and (ii) specifically for an upper tail that is unequivocally characterized by a Pareto distribution. This should show whether Zipf's law is confirmed (with a coefficient of -1$\pm$-0.1) or not and how strong coefficients differ. Given the often found larger variation of Zipf's law, the application of a relaxed definition (e.g. -1$\pm$-0.3) could be considered, but in the end, any such definition of intervals remains arbitrary. 

\hspace*{3em}As a further important intermediate step of the analysis, we explore spatial autocorrelation along the city ranks by estimating local Moran-I coefficients (LISA). This helps to immediately unveil the sections of the distribution, where spatial dependence occurs and where the mean of LISA coefficients is significantly different from zero. The aim is to identify an upper tail that is Pareto-distributed but perturbed by spatial autocorrelation, as indicated by a coefficient with a significant minor or major deviation from a distribution that corresponds to a (non-spatial) Zipf law. If the coefficient were disturbed by spatial dependence, we could conclude that a non-spatial analysis of Zipf's law is insufficient. This is because for two countries with an identical Pareto coefficient, significant spatial dependence for only one country would imply a bias. 

\hspace*{3em} How do we proceed? We first truncate the observations at the city rank with the cumulative 50 percent of the entire urban population. This is also highlighted by Malevergne et al. (2011) who apply an UMPU test to confirm a hybrid rank-size distribution of cities. We can thus well assume that these upper sections with 50 percent of the population belong to a Pareto shaped tail, but the remaining number of observations after truncation is often rather small. In other words, there is a notable variation in kurtosis: For Australia, already the first three out of 274 ranks of the city distribution represent 1/2 of the entire city population, while for Italy, it takes 159 out of 1,356 ranks to display 50 percent of the urban population. Furthermore, restricting the analysis on few upper observations may conceal important spatial effects from cities below the cut-off. It is possible that city ranks below the truncation point (potentially exhibiting local spatial autocorrelation) still belong to the Pareto tail. Therefore, the different "conspicuous" sections of the rank-size distributions below the truncation point have been then tested for log-normality by moving windows. This was done by (i) visual inspection of the histograms and (ii) a Shapiro-Wilk test of the log-transformed observations of S. Furthermore, we back-up the result by generating fitted log-normal and Pareto distributions from the observed data in the different sections of the size distribution. These alternatively fitted data are then compared to the observed data by calculating the Kullback-Leibler divergence to confirm that the Pareto distribution is more probable than a log-normal one. This idea was \textit{inter alia} posited by Bee et al.(2011). The approach of that relative entropy test to recognize the respectively more adequate distribution type is defined by:

\begin{equation}
    D_{ KL }(p\parallel q)=\sum_{ x \in X } { p(x_{ i } )} log\frac{ p(x_{ i }) }{ q(x_{ i } )}
\end{equation}
where p and q represent the probabilities of observations and the respectively fitted distributions, in this case a Pareto versus an alternative lognormal type in a rank range within X. 

\hspace*{3em}The advantage of the World Cities Database constitutes its actuality, the large number of observations (around 48,000), the ready availability of geo-coordinates and the definition of size in terms of population in the larger urban zone of a city. Evidence has suggested that such functional city sizes - rather than strict administrative boundaries - better represent Zipf’s law (Budde and Neumann 2019).
\section{Results}
Preliminary results of our empirical exploration show a differentiated picture among different dimensions. For some countries, there are major differences in the Pareto coefficient when testing the full versus the truncated dataset, for other countries coefficients remain similar. At the same time there are some countries with significant spatial dependence only for the full dataset, some with minor (e.g. USA), others with moderate influence on the Pareto coefficient (e.g. France), but there are also countries where spatial dependence occurs in both, the full as well as the truncated dataset (e.g. Sweden). 
\begin{figure}
    \centering
    \includegraphics[width=0.75\linewidth]{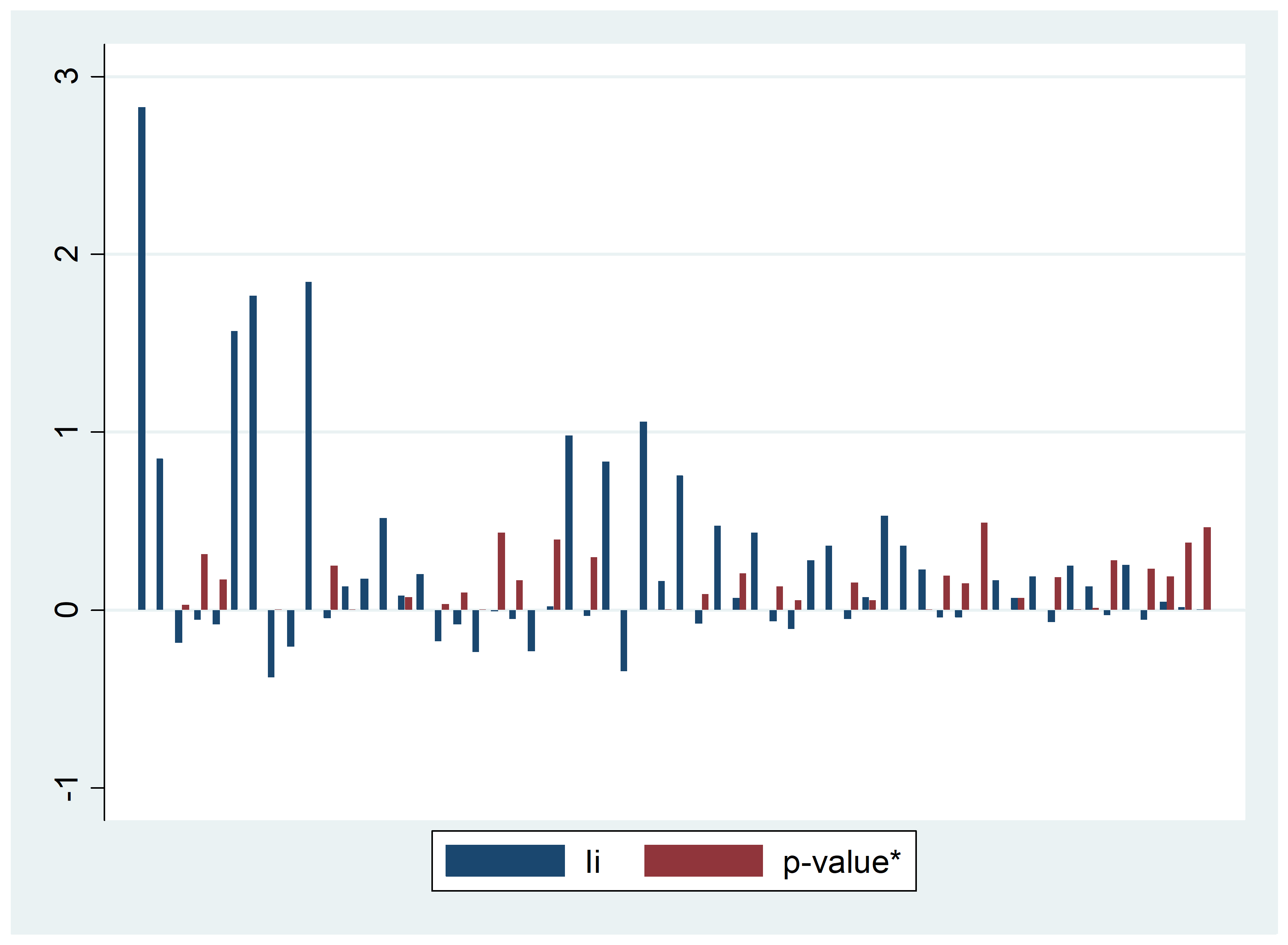}
    \caption{Belgium: Local Moran I coefficients (LISA) and related p-values for the upper 57 observations}
    \label{fig:enter-label}
\end{figure}
It appears that spatial dependence influences are more likely in the larger and medium OECD countries (USA, Germany, France, Italy, Canada, Sweden, Belgium) while this is less the case in rather small countries (except Estonia). Perhaps in small countries there is not so much difference between models with and without controls for distance (because there is not so much spatial distance) but evidence is not fully conclusive. For those countries where a spatial dependence effect could be established, we can report the following deviations of the spatial Pareto coefficient from its non-spatial counterpart: Canada (-.001; SAR full range), Chile (-.034; SAR full range), Netherlands (-.001; SAR full range), France (-.01; SEM Pareto range), Italy (.005; SAR Pareto range), Germany (-.016; SAR Pareto range), USA (.002; SEM Pareto range) and Sweden (-.012; SEM full range and -.044; SAR Pareto range). The regression tables (Appendix) show the two Pareto coefficients (spatial and non-spatial) for the entire range and for the truncated sample. Only countries with a significant distance effect are presented. As an example, we demonstrate the approach for Belgium, a medium-sized OECD country. The rank of 50 percent of cumulative population is the 57th largest Belgian city (Chatelet). The full range of Belgian cities displayed by the dataset is 397.  With coefficients smaller than -1.1, Zipf's law cannot be confirmed for Belgium, neither for the full range of cities nor for the upper tail (with a relaxed definition of Zipf, the law would still hold for the truncated distribution, but only for the non-spatial regression; spatial dependence effects move the coefficient outside the above defined Zipf range). The Shapiro-Wilk test confirms non-normality of the log-transformed population data for the 57 observations of the upper tail (W=.758; p=.000). The back-up analysis with Kullback-Leibler divergence, estimated by the Stata command "reldist divergence" shows a significantly lower value when matching the observations with a fitted Pareto distribution as compared to matching them with an alternatively fitted log-normal distribution: KL(Pareto)=.31; KL(log-normal)=.90. We conclude that the two different tests executed clearly confirm a Pareto shape for the explored upper tail. 
\\
\hspace*{3em}In addition to the spatially augmented Pareto regressions we also inspect local spatial autocorrelation along the city ranks in order to detect those sections of the respective distributions where significant local spatial autocorrelation occurs (cf. Figure 1; where no or a red bar smaller than 0.1 occurs). For Belgium, the density of significant spatial autocorrelation is substantially higher in the upper tail, although there is evidence of occult occasional spatial dependence also among lower city ranks (not displayed). Here, it is also possible to identify the city sections from which a disturbing effect of distance on the Pareto coefficient might arise. However, spatial dependence in the lower tail is less relevant for explaining spatially induced deviations from Zipf's law, at least if for the lower tail the null hypothesis of log-normality cannot be rejected. 
\\
\hspace*{3em}Collectively, the upper Pareto tail of the rank-size distribution of Belgian cities represents the major part of cities with significant positive or negative spatial autocorrelation. The example of Belgium shows that the existence of spatial dependence may have a significant influence on the Pareto coefficient of the rank-size distribution of cities. Similar findings can be reported for France, Germany and Sweden, to a lesser extent also for Italy and the USA (cf. Appendix). The economic explanation for this peculiarly varying degree of spatial autocorrelation in the rank-size distribution of cities among the OECD countries requires further research.

\section{References:}

\hspace{1.5em}Arcaute E, Ramasco JJ (2022) Recent advances in urban system science: Models and data. PLoS ONE 17(8): e0272863

Bee M, Riccaboni M, Schiavo S (2011) Pareto versus lognormal: A maximum entropy test. Phys Rev E 84: 026104

Bergs R (2021) Spatial dependence in the rank-size distribution of cities – weak but not negligible. PLoS ONE 16(2): e0246796

Budde R, Neumann U (2019) The size ranking of cities in Germany: caught by a MAUP?. GeoJournal 84, 1447–1464.

Gabaix X, Ibragimov R (2011) Rank - 1/2: A simple way to improve the OLS estimations of tail exponents. J Bus Econ Stat 29(1)

Gan L, Li D, Song S (2006) Is the Zipf law spurious in explaining city-size distributions? Econ Lett 92)2

Griffiths D (2022) The United States Urban Hierarchy: An Update. Front. Sustain. Cities 4

Hsu WT (2011) Central place theory and city size distribution. Econ J 122(563) 

Le Gallo J, Chasco C (2008) Spatial analysis of urban growth in Spain, 1900–2001. Empir Econ 34: 59-80

Malevergne Y, Pisarenko V, Sornette D (2011) Testing the Pareto against the lognormal distributions with the uniformly most powerful unbiased test applied to the distribution of cities. Phys Rev E 83: 036111

Xiao Y, Gong P (2022) Removing spatial autocorrelation in urban scaling analysis. Cities 124: 103600.

Zipf GK (1932) Selected studies on the principle of relative frequency in language. Cambridge, MA: Harvard University Press

\section{Data:}

NGIA, US Geological Survey, US Census Bureau, and NASA (2024) World Cities Database. Link: https://simplemaps.com/data/world-cities

\newpage
\section{Appendix: OECD countries with significant Spatial dependence in the rank-size distribution of cities: (Table 1: Truncated range of cities; Table 2: For information only: Full range of cities)}
\begin{table}[h]
\tiny
\begin{tabular}{lllllllllllllllll}
Belgium      & LN R      & lambda    & LN R      & rho      & LN R (non spat.)      & Shapiro-Wilk & Kullback-Leibler &  &  &  &  &  &  &  &  &  \\
LN S         & -1.343*** &           & -1.296*** &          & -1.293*** & ***            & yes              &  &  &  &  &  &  &  &  &  \\
Constant     & 16.82***  & -2.644*** & 16.07***  & 0.0951   & 16.28***  &              &                  &  &  &  &  &  &  &  &  &  \\
Observations & 57        & 57        & 57        & 57       & 57        &              &                  &  &  &  &  &  &  &  &  &  \\
R-squared    &           &           &           &          & 0.963     &              &                  &  &  &  &  &  &  &  &  &  \\
             &           &           &           &          &           &              &                  &  &  &  &  &  &  &  &  &  \\
France       & LN R      & lambda    & LN R      & rho      & LN R (non spat.)      & Shapiro-Wilk & Kullback-Leibler &  &  &  &  &  &  &  &  &  \\
LN S         & -1.072*** &           & -1.089*** &          & -1.082*** & ***            & yes              &  &  &  &  &  &  &  &  &  \\
Constant     & 14.86***  & 0.501**   & 14.52***  & 0.205*   & 15.02***  &              &                  &  &  &  &  &  &  &  &  &  \\
Observations & 66        & 66        & 66        & 66       & 66        &              &                  &  &  &  &  &  &  &  &  &  \\
R-squared    &           &           &           &          & 0.903     &              &                  &  &  &  &  &  &  &  &  &  \\
             &           &           &           &          &           &              &                  &  &  &  &  &  &  &  &  &  \\
Germany      & LN R      & lambda    & LN R      & rho      & LN R (non spat.)      & Shapiro-Wilk & Kullback-Leibler &  &  &  &  &  &  &  &  &  \\
LN S         & -1.108*** &           & -1.094*** &          & -1.110*** & ***            & yes              &  &  &  &  &  &  &  &  &  \\
Constant     & 16.08***  & -0.498    & 16.88***  & -0.309** & 16.11***  &              &                  &  &  &  &  &  &  &  &  &  \\
Observations & 98        & 98        & 98        & 98       & 98        &              &                  &  &  &  &  &  &  &  &  &  \\
R-squared    &           &           &           &          & 0.987     &              &                  &  &  &  &  &  &  &  &  &  \\
             &           &           &           &          &           &              &                  &  &  &  &  &  &  &  &  &  \\
Italy        & LN R      & lambda    & LN R      & rho      & LN R (non spat.)      & Shapiro-Wilk & Kullback-Leibler &  &  &  &  &  &  &  &  &  \\
LN S         & -1.367*** &           & -1.374*** &          & -1.369*** & ***            & yes              &  &  &  &  &  &  &  &  &  \\
Constant     & 18.48***  & 0.416     & 18.13***  & 0.119**  & 18.51***  &              &                  &  &  &  &  &  &  &  &  &  \\
Observations & 159       & 159       & 159       & 159      & 159       &              &                  &  &  &  &  &  &  &  &  &  \\
R-squared    &           &           &           &          & 0.987     &              &                  &  &  &  &  &  &  &  &  &  \\
             &           &           &           &          &           &              &                  &  &  &  &  &  &  &  &  &  \\
Sweden       & LN R      & lambda    & LN R      & rho      & LN R (non spat.)      & Shapiro-Wilk & Kullback-Leibler &  &  &  &  &  &  &  &  &  \\
LN S         & -0.728*** &           & -0.685*** &          & -0.729*** & ***            & yes              &  &  &  &  &  &  &  &  &  \\
Constant     & 9.609***  & -0.022    & 9.743***  & -0.579** & 9.621***  &              &                  &  &  &  &  &  &  &  &  &  \\
Observations & 10        & 10        & 10        & 10       & 10        &              &                  &  &  &  &  &  &  &  &  &  \\
R-squared    &           &           &           &          & 0.922     &              &                  &  &  &  &  &  &  &  &  &  \\
             &           &           &           &          &           &              &                  &  &  &  &  &  &  &  &  &  \\
USA          & LN R      & lambda    & LN R      & rho      & LN R (non spat.)      & Shapiro-Wilk & Kullback-Leibler &  &  &  &  &  &  &  &  &  \\
LN S         & -1.078*** &           & -1.075*** &          & -1.076*** & ***            & yes              &  &  &  &  &  &  &  &  &  \\
Constant     & 17.70***  & -0.658*   & 17.73***  & -0.0233  & 17.67***  &              &                  &  &  &  &  &  &  &  &  &  \\
Observations & 101       & 101       & 101       & 101      & 101       &              &                  &  &  &  &  &  &  &  &  &  \\
R-squared    &           &           &           &          & 0.968     &              &                  &  &  &  &  &  &  &  &  & 
\end{tabular}
\caption{OECD countries with significant spatial dependence in the truncated sample; Kullback-Leibler (yes) confirms a divergence for a Pareto distribution lower than for a log-normal one; Significance ***=0.01 level. **=0.05 level, *=0.1 level}
\label{tab:my-table}
\end{table}
\begin{table}[]
\tiny
\begin{tabular}{llllllllll}
            &                         & LN R      & lambda    & LN R      & rho       & LN R (non spat.)     &  &  &  \\
Australia   & LN S                    & -0.451*** &           & -0.470*** &           & -0.451*** &  &  &  \\
            & Constant                & 7.893***  & -0.00448  & 9.869***  & -0.452*** & 7.893***  &  &  &  \\
            & Observations            & 274       & 274       & 274       & 274       & 274       &  &  &  \\
            & R-squared               &           &           &           &           & 0.823     &  &  &  \\
Belgium     & LN S                    & -1.469*** &           & -1.462*** &           & -1.463*** &  &  &  \\
            & Constant                & 18.25***  & -3.939*** & 18.13***  & 0.0141    & 18.20***  &  &  &  \\
            & Observations            & 397       & 397       & 397       & 397       & 397       &  &  &  \\
            & R-squared               &           &           &           &           & 0.986     &  &  &  \\
Canada      & LN S                    & -0.886*** &           & -0.888*** &           & -0.887*** &  &  &  \\
            & Constant                & 13.30***  & -0.0937   & 13.41***  & -0.0222*  & 13.30***  &  &  &  \\
            & Observations            & 485       & 485       & 485       & 485       & 485       &  &  &  \\
            & R-squared               &           &           &           &           & 0.997     &  &  &  \\
Chile       & LN S                    & -0.479*** &           & -0.469*** &           & -0.503*** &  &  &  \\
            & Constant                & 8.650***  & 0.752***  & 6.437***  & 0.537***  & 8.897***  &  &  &  \\
            & Observations            & 253       & 253       & 253       & 253       & 253       &  &  &  \\
            & R-squared               &           &           &           &           & 0.643     &  &  &  \\
Estonia     & LN S                    & -0.647*** &           & -0.634*** &           & -0.664*** &  &  &  \\
            & Constant                & 7.580***  & -1.575**  & 8.108***  & -0.406    & 7.733***  &  &  &  \\
            & Observations            & 20        & 20        & 20        & 20        & 20        &  &  &  \\
            & R-squared               &           &           &           &           & 0.95      &  &  &  \\
Netherlands & LN S                    & -1.116*** &           & -1.116*** &           & -1.117*** &  &  &  \\
            & Constant                & 15.25***  & -0.512    & 16.41***  & -0.262**  & 15.27***  &  &  &  \\
            & Observations            & 361       & 361       & 361       & 361       & 361       &  &  &  \\
            & R-squared               &           &           &           &           & 0.97      &  &  &  \\
New Zealand & LN S                    & -0.497*** &           & -0.501*** &           & -0.504*** &  &  &  \\
            & Constant                & 7.260***  & 0.603**   & 6.561***  & 0.289*    & 7.312***  &  &  &  \\
            & Observations            & 58        & 58        & 58        & 58        & 58        &  &  &  \\
            & R-squared               &           &           &           &           & 0.841     &  &  &  \\
Norway      & LN S                    & -1.043*** &           & -1.041*** &           & -1.037*** &  &  &  \\
            & Constant                & 13.12***  & 0.435     & 13.57***  & -0.144*   & 13.07***  &  &  &  \\
            & Observations            & 124       & 124       & 124       & 124       & 124       &  &  &  \\
            & R-squared               &           &           &           &           & 0.957     &  &  &  \\
Poland      & LN S                    & -1.085*** &           & -1.086*** &           & -1.087*** &  &  &  \\
            & Constant                & 15.06***  & -0.739    & 15.43***  & -0.0774** & 15.07***  &  &  &  \\
            & Observations            & 465       & 465       & 465       & 465       & 465       &  &  &  \\
            & R-squared               &           &           &           &           & 0.994     &  &  &  \\
Slovakia    & LN S                    & -1.195*** &           & -1.171*** &           & -1.178*** &  &  &  \\
            & Constant                & 14.25***  & -1.576*** & 13.77***  & 0.0881*   & 14.09***  &  &  &  \\
            & Observations            & 83        & 83        & 83        & 83        & 83        &  &  &  \\
            & R-squared               &           &           &           &           & 0.987     &  &  &  \\
Sweden      & LN S                    & -1.015*** &           & -1.024*** &           & -1.027*** &  &  &  \\
            & Constant                & 13.27***  & 0.906***  & 12.39***  & 0.267***  & 13.35***  &  &  &  \\
            & Observations            & 150       & 150       & 150       & 150       & 150       &  &  &  \\
            & R-squared               &           &           &           &           & 0.985     &  &  &  \\

\end{tabular}
\caption{For information only: OECD countries with significant spatial dependence in the rank-size distribution of cities (full range of cities); Significance: ***=0.01 level, **=0.05 level, *=0.1 level. Some larger countries (Colombia, France, Germany, Italy, Japan, Mexico, Spain, Turkey, UK and USA) could not be analyzed due to the size of the weight matrix and limits of computing capacity}
\label{tab:my-table}
\end{table}

\end{document}